\documentclass[preprint,preprintnumbers,showpacs,aps,prd,amssymb]{revtex4-1}

\usepackage{graphicx}
\usepackage{bm}
\usepackage{amsmath}

\def\calC{{\cal C}}

\def\calH{{\cal H}}
\def\calL{{\cal L}}
\def\calO{{\cal O}}

\def\calU{{\cal U}}

\def\Bbar{{\bar B}}

\def\cbar{{\bar c}}
\def\hbar{{\bar h}}

\def\Ds{D^{(*)}}
\def\dU{{d_\calU}}
\def\LU{\Lambda_\calU}
\def\eU{{\epsilon_\calU}}

\def\SM{{\rm SM}}
\def\Br{{\rm Br}}

\def\nn{\nonumber}

\begin{document}
\title{$B$ anomalies with unparticles}
\author{Jong-Phil Lee}
\email{jongphil7@gmail.com}
\affiliation{Sang-Huh College,
Konkuk University, Seoul 05029, Korea}

\begin{abstract}
We analyze the $B$ anomalies associated with the $B\to\Ds\tau\nu$ decays in the unparticle model. 
The fraction of the branching ratios $R(\Ds)$ and other parameters related to the polarization are fitted
to the experimental data by minimizing $\chi^2$.
The best-fit values are $R(D)_{\rm best}=0.371$ and $R(D^*)_{\rm best}=0.266$
which are still larger than the standard model predictions.
We find that our results safely render the branching ratio $\Br(B_c\to\tau\nu)$ below $10\%$.
\end{abstract}

\maketitle
\section{Introduction}
Flavor physics plays an important role in particle physics to probe new physics (NP) as well as to test the standard model (SM).
The SM has been very successful to describe the nature so far,
and no explicit evidence for the NP has been observed yet.
But the SM is incomplete in many respects and we anticipate the appearance of the NP.
Recently some anomalies appear in $B$ physics.
Especially the fraction of ($\ell=e,\mu$)
\begin{equation}
R(\Ds)\equiv\frac{{\rm Br}(B\to \Ds\tau\nu)}{{\rm Br}(B\to \Ds\ell\nu)}~,
\end{equation}
shows a tension with the SM predictions \cite{Amhis}
\begin{eqnarray}
R(D)_{\rm SM} &=&0.299\pm0.003~,\nonumber\\
R(D^*)_{\rm SM} &=&0.258\pm0.005~.
\label{RDSM}
\end{eqnarray}
Experimental data up to now favor larger values of the fraction \cite{BaBar_PRL,BaBar1,Belle1,Belle1607,Belle1703,Belle1612,Belle1709,Belle1904,LHCb1,LHCb2}.
The world averages by the Heavy Flavor Averaging Group(HFLAV) Collaboration are \cite{HFAG2019} 
\begin{eqnarray}
R(D)_{\rm HFLAV} &=& 0.340\pm0.027\pm0.013~,\nn\\
R(D^*)_{\rm HFLAV} &=& 0.295\pm0.011\pm0.008~,
\label{RDHFLAV}
\end{eqnarray}
which exceed the SM predictions by $3.08\sigma$.
The discrepancy shows the lepton universality violation. 
If we allow anomalous $\tau$ couplings then one can easily enhance the $R(\Ds)$ values, 
as in \cite{JPL_PRD96}.
There have been many attempts to explain the discrepancy in various NP models,
like leptoquark models  \cite{Dorsner,Alonso,Bauer,Barbieri,DiLuzio,Calibbi,Becirevic}, 
composite models \cite{Barbieri2,Buttazzo,Bordone,Matsuzaki}, 
extra dimensions \cite{Megias1,Megias2,DAmbrosio,Blanke0,JPL_nmUED}, 
etc. \cite{Kang,Huang,Bardhan}.
\par
In addition to $R(\Ds)$, there are other observables related to the polarizations 
in $B\to\Ds\tau\nu$ decays., involving $\tau$ as well as $D^*$ polarizations.
Firstly, the polarization asymmetry of $\tau$ is defined as
\begin{equation}
P_\tau(\Ds) \equiv \frac{\Gamma^{D^{(*)}}_\tau(+)-\Gamma^{D^{(*)}}_\tau(-)}
                                          {\Gamma^{D^{(*)}}_\tau(+)+\Gamma^{D^{(*)}}_\tau(-)}~,
\end{equation}
where $\Gamma^{\Ds}_\tau(\pm)$ is the decay width corresponding to $(\pm)~\tau$ helicity. 
Expected values from the SM are \cite{Tanaka2010,Tanaka2012}
\begin{equation}
P_\tau(D)_{\rm SM} = 0.325\pm0.009~,~~~
P_\tau(D^*)_{\rm SM} = -0.497\pm0.013~.
\end{equation}
Experimentally the measured value is \cite{Belle1612,Belle1709}
\begin{equation}
P_\tau(D^*) = -0.38\pm0.51^{+0.21}_{-0.16}~.
\end{equation}
Secondly, the longitudinal $D^*$ polarization is given by
\begin{equation}
F_L(D^*)\equiv\frac{\Gamma(B\to D^*_L\tau\nu)}{\Gamma(B\to D^*\tau\nu)}~.
\end{equation}
The SM prediction is \cite{Alok2016}
\begin{equation}
F_L(D^*)_{\rm SM} = 0.46\pm0.04~,
\end{equation}
which is smaller than the Belle's measurement \cite{Abdesselam}
\begin{equation}
F_L(D^*)  = 0.60\pm0.08\pm0.035~.
\end{equation}
\par
Among other NP models, the unparticle($\calU$) scenario is quite interesting and unique \cite{Georgi,unitarity}.
Unparticles are low-energy realization of a scale-invariant hidden sector at some high-energy scale.
In the context of the effective field theory, unparticles behave like a fractional number of particles.
Usually unparticles contribute in the form of $\lambda_\calU^2\left(M_{\rm EW}^2/\LU^2\right)^{\dU}$,
where $M_{\rm EW}$ is the electroweak scale, $\lambda_\calU$ is some relevant coupling, 
$\LU$ is a new high-energy scale for the scale invariance, 
and $\dU$ is the scaling dimension of the unparticle operators.
Typical NP involves ordinary particles with a definite integer of $\dU$
while for unparticles $\dU$ is a free parameter, which makes the scenario more interesting.
One can suppress or enhance the NP effects by not only the new scale $\LU$ but also $\dU$.
In general scalar and vector unparticles can contribute together. 
But as for the vector unparticles the lower bound of the associated scaling dimension is larger than 
that for the scalar unparticles, resulting in more suppressions of $\left(M_{\rm EW}^2/\LU^2\right)$. 
Unparticles can affect the $B$ physics in many ways \cite{Geng,Mohanta1,JPL_RD_U},
for example, $B_s$-${\bar B}_s$ mixing \cite{Lenz,Mohanta2,Parry,JPL_BsBsbar}
(for meson mixing, see \cite{,Li,Chen}),
and $B_s\to\mu^+\mu^-$ \cite{JPL_Bs2mumu}, etc.
These are included as constraints in this analysis.
\par
The paper is organized as follows.
In the next section, unparticle descriptions are given for the relevant observables.
Our results and discussions appear in Sec.\ III.
We conclude in Sec.\ IV.
%
\section{Unparticles and Observables}
The relevant Lagrangian involving scalar unparticles $\calO_\calU$ and vector ones $\calO_\calU^\mu$ is given by
\cite{JPL_RD_U,JPL_Bs2mumu}
\begin{equation}
 \calL_\calU
=\sum_f\left[
\frac{c_S^{f'f}}{\Lambda_\calU^\dU}{\bar f'}\gamma_\mu(1-\gamma_5)f~\partial^\mu\calO_\calU
  +\frac{c_V^{f'f}}{\Lambda_\calU^{d_V-1}}{\bar f'}\gamma_\mu(1-\gamma_5)f~\calO_\calU^\mu\right]~,
\label{LU}
\end{equation}
where $c_{S,V}^{L(R), f'f}$ are dimensionless couplings, and $\dU(d_V)$ is the scaling dimension of $\calO_\calU(\calO_\calU^\mu)$.
Note that we only consider the left-handed currents for simplicity.
The unitarity constraint requires that $\dU\ge 1$ and $d_V\ge 3$ \cite{unitarity}.
Typically the scalar and vector contributions amount to 
$\sim(M_{\rm EW}^2/\LU^2)^\dU$ and $\sim(M_{\rm EW}^2/\LU^2)^{d_V-1}$, respectively.
One can expect that effects of vector unparticles are very suppressed compared to 
those of scalar ones due to the unitarity constraints \cite{JPL_Bs2mumu}.
In this analysis we put for simplicity $\dU=1+\eU$ and $d_V=3+\eU$ with $0\le\eU\le 1$.
\par
The effective Hamiltonian for $b\to c\ell\nu$ with unparticles is 
\begin{equation}
\calH_{\rm eff}
=\frac{4G_F}{\sqrt{2}}V_{cb}\sum_{\ell=\mu,\tau}\left[
(1+C_V^{L,\ell})\calO_V^{L,\ell} + C_S^{L,\ell}\calO_S^{L,\ell}+C_S^{R,\ell}\calO_S^{R,\ell}\right]~,
\end{equation}
where the operators $\calO_{V,S}^\ell$ are defined by
\begin{eqnarray}
\calO_V^{L,\ell} &=& \left(\cbar_L\gamma^\mu b_L\right)\left({\bar\ell}_L\gamma_\mu\nu_{\ell L}\right)~,\\
\calO_S^{L,\ell} &=& \left(\cbar_R b_L\right) \left({\bar\ell}_R\nu_{\ell L}\right)~,\\
\calO_S^{R,\ell} &=& \left(\cbar_L b_R\right) \left({\bar\ell}_R\nu_{\ell L}\right)~.
\end{eqnarray}
The Wilson coefficients are (putting $c_{S,V}^{cb}\equiv c_{S,V}^q$ and $c_{S,V}^{\ell\nu}\equiv c_{S,V}^\ell$) 
\begin{eqnarray}
2\sqrt{2}G_F m_B^2 V_{cb} C_V^{L,\ell}&=&
\frac{-A_{d_V}e^{-i\eU\pi}}{2\sin\eU\pi}\left(\frac{m_B}{\LU}\right)^{2\eU+4}c_V^q c_V^\ell~,\\
2\sqrt{2}G_F m_B^2 V_{cb} C_S^{L,\ell}&=&
\frac{-A_{d_S}e^{-i\eU\pi}}{2\sin\eU\pi}\left(\frac{m_B}{\LU}\right)^{2\eU+2}\left(\frac{m_\ell m_c}{m_B^2}\right)c_S^q c_S^\ell~,\\
2\sqrt{2}G_F m_B^2 V_{cb} C_S^{R,\ell}&=&
\frac{A_{d_S}e^{-i\eU\pi}}{2\sin\eU\pi}\left(\frac{m_B}{\LU}\right)^{2\eU+2}\left(\frac{m_\ell m_b}{m_B^2}\right)c_S^q c_S^\ell~,
\end{eqnarray}
where
\begin{eqnarray}
\label{AdU}
A_\dU&=&\frac{16\pi^{5/2}}{(2\pi)^{2\dU}}
		\frac{\Gamma(\dU+1/2)}{\Gamma(\dU-1)\Gamma(2\dU)}~,\\
\label{phidU}		
\phi_\calU&=&(\dU-2)\pi~.
\end{eqnarray}
Here we have neglected the terms of $\calO(m_B^2/\LU^2)$ in $C_S^{L,R}$.
They involve vector couplings of $c_V^q c_V^\ell$.
Note that the couplings appear in combined forms of $c_{S,V}^qc_{S,V}^\ell$.
%
Numerically the observables for $B\to\Ds\ell\nu_\ell$ decays are (at $\mu=m_b$ scale) \cite{Blanke,Aoki,Bernlochner}
\begin{equation}
\label{obs1}
R\left(\Ds\right)=R_\SM\left(\Ds\right)\frac{2\left[1+r_{\tau,\calU}^{(*)}\right]}{1+\left[1+r_{\mu,\calU}^{(*)}\right]}~,
\end{equation}
where 
\begin{eqnarray}
1+r_{\ell,\calU} &=& \left|1+C_V^{L,\ell}\right|^2 
     + 1.54{\rm Re}\left[\left(1+C_V^{L,\ell}\right)\left(C_S^{L,\ell}+C_S^{R,\ell}\right)^*\right]
     + 1.09\left|C_S^{L,\ell}+C_S^{R,\ell}\right|^2     ~,\nn\\
1+r^*_{\ell,\calU} &=& \left|1+C_V^{L,\ell}\right|^2 
      + 0.13{\rm Re}\left[\left(1+C_V^{L,\ell}\right)\left(C_S^{R,\ell}-C_S^{L,\ell}\right)^*\right]
     + 0.05\left|C_S^{R,\ell}-C_S^{L,\ell}\right|^2 
\end{eqnarray}
and
\begin{eqnarray}
P_\tau(D) &=& 
1 - 0.68 \frac{\left|1+C_V^{L,\tau}\right|^2}{1+r_{\tau,\calU}}~,\\
P_\tau(D^*) &=&
1 - 1.49 \frac{\left|1+C_V^{L,\tau}\right|^2}{1+r^*_{\tau,\calU}}~,\\
F_L(D^*) &=&
1 -0.54\frac{\left|1+C_V^{L,\tau}\right|^2}{1+r^*_{\tau,\calU}}~.
\label{obs5}
\end{eqnarray}
%
\begin{table}
\begin{tabular}{|c|| cc |}\hline
              & ~$R(D)$ & ~$R(D^*)$   \\\hline
 BABAR & ~$0.440\pm0.058\pm0.042$ & ~$0.332\pm0.024\pm0.018$ \cite{BaBar1} \\
 Belle(2015) & ~$0.375\pm0.064\pm0.026$ & ~$0.293\pm0.038\pm0.015$ \cite{Belle1} \\
 Belle(2016) & ~$-$ & ~$0.302\pm0.030\pm0.011$ \cite{Belle1607} \\
 Belle(2017) & ~$-$ & ~$0.276\pm0.034^{+0.029}_{-0.026}$ \cite{Belle1703} \\
 Belle(2017) & ~$-$ & ~$0.270\pm0.035^{+0.028}_{-0.025}$ \cite{Belle1612,Belle1709} \\
 Belle(2019) & ~$0.307\pm0.037\pm0.016$ & ~$0.283\pm0.018\pm0.014$ \cite{Belle1904}\\
 LHCb(2015) & ~$-$ & ~$0.336\pm0.027\pm0.030$ \cite{LHCb1} \\
 LHCb(2017) & ~$-$ & ~$0.291\pm0.019\pm0.026\pm0.013$ \cite{LHCb2} \\\hline\hline
 & $P_\tau(D^*)$ & $F_L(D^*)$ \\\hline
 Belle(2017) & $-0.38\pm0.51^{+0.21}_{-0.16}$\cite{Belle1612,Belle1709} & $-$ \\
 Belle(2019)  & $-$ & $0.60\pm0.08\pm0.04$ \cite{Abdesselam} \\
 \hline
 \end{tabular}
\caption{Summary of experimental data for $R(\Ds)$, $P_\tau(\Ds)$ and $F_L(D^*)$.
The uncertainties are $\pm$(statistical)$\pm$(systematic).
The correlations between $R(D)$ and $R(D^*)$ for BABAR, Belle(2015), and Belle(2019) results are
$-0.31$, $-0.50$ and $-0.51$, respectively \cite{HFAG2019}.}
\label{T1}
\end{table}
%
\section{Results and Discussions}
The experimental data are summarized in Table \ref{T1} \cite{JPL_nmUED}.
For the uncertainties of LHCb(2017), see discussion in \cite{LHCb2}.
The $\chi^2$ is defined by
\begin{equation}
\chi^2\equiv\sum_{i,j}
\left[ \calO_i^{\rm exp}-\calO_i^{\rm th}\right]\calC_{ij}^{-1}
\left[\calO_j^{\rm exp}-\calO_j^{\rm th}\right]~,
\label{chisq}
\end{equation}
where $\calO_i^{\rm exp}$ are the experimental data in Table \ref{T1}
and $\calO_i^{\rm th}$ are the theoretical calculations from Eqs.(\ref{obs1})-(\ref{obs5}).
Here $\calC_{ij}$ are the correlation matrix elements.
There are relevant constraints from flavor physics.
We include $B_s$-${\bar B_s}$ mixing\cite{Lenz,MsSM,Luzio2018}, 
$B\to D^+D^-$ decay\cite{Zwicky_DD}, 
 $B_s\to\mu^+\mu^-$ decay, and $B_c\to\tau\nu$ decay.
 The first two involve only quark couplings $c_{S,V}^q$ while others do $c_{S,V}^q c_{S,V}^\ell$.
 For these constraints we only consider the scalar couplings for simplicity since they are dominant.
%
For $B_s$-$\Bbar_s$ mixing, the mass difference $\Delta M_s$ is given by
\begin{equation}
\Delta M_s =2|M_{12}^\SM||\Delta_s|=\Delta M_s^\SM|\Delta_s|
\end{equation}
where 
\begin{equation}
M_{12}^\SM=
\frac{G_F^2 m_W^2}{12\pi^2}\left(V_{ts}^*V_{tb}\right)^2 m_{B_s}B_{B_s}f_{B_s}^2\eta_{B_s}S_0(x_t)~.
\end{equation}
Here $S_0(x_t\equiv m_t^2/m_W^2)$ is the Inami-Lim function, 
and other hadronic parameters can be found in \cite{Lenz,JPL_BsBsbar}.
And $\Delta_s$ is defined by
\begin{equation}
\Delta_s = 1+\left(c_S^q\right)^2 f_s(\dU)\cot(\dU\pi)-i\left(c_S^q\right)^2 f_s(\dU)~,
\end{equation}
where
\begin{equation}
f_s(\dU)\equiv
\frac{1}{M_{12}^\SM}\frac{2\pi^{5/2}}{(2\pi)^{2\dU}}
\frac{\Gamma(\dU+1/2)}{\Gamma(\dU-1)\Gamma(2\dU)}
\left( \frac{f_{B_s}^2}{m_{B_s}} \right)
\left( \frac{m_{B_s}^2}{\LU^2} \right)^\dU
\frac{5}{3}~.
\end{equation}
The decay rate difference is \cite{Lenz} 
\begin{equation}
\Delta\Gamma_s=2|\Gamma_{12,s}|\cos(\phi^\SM_s+\phi^\Delta_s)
=(0.096\pm0.039)~{\rm ps}^{-1}\times\cos(\phi^\SM_s+\phi^\Delta_s)~,
\end{equation}
where $\phi^{\SM(\Delta)}_s$ is the phase of the SM($\Delta_s$) contribution.
Another observable is the CP asymmetry parameter \cite{Lenz}
\begin{equation}
a_{fs}^s = {\rm Im}\frac{\Gamma_{12,s}}{M_{12}}~.
\end{equation}
We use the following experimental data \cite{HFAG_Bsmix}
\begin{eqnarray}
\Delta M_s &=&17.757\pm0.021~{\rm ps}^{-1}~,\\
\Delta\Gamma_s &=& 0.090\pm0.005~{\rm ps}^{-1}~,\\
a_{fs}^s &=& (0.6\pm 2.8)\times 10^{-3}~.
\end{eqnarray}
%
\par
As for $B\to DD$ decays, the branching ratio is \cite{Zwicky_DD}
\begin{equation}
{\rm Br}(B\to D^+D^-)={\rm Br}_{DD}^\SM f_{\Delta_{DD}}~,
\end{equation}
where ${\rm Br}_{DD}^\SM$ is the SM contribution, and
\begin{eqnarray}
f_{\Delta_{DD}} &=& \left| 1+\Delta_{DD} e^{-i\phi_\calU} e^{-i\phi_w} \right|~,\\
\Delta_{DD} &=&
\frac{|c_S^q|^2}{a_1|V_{cb}V_{cd}|}
\frac{A_\dU}{2\sin(\dU\pi)}
\left(\frac{m_b+m_c}{m_c}\right)
\frac{\sqrt{2}}{G_F\LU^2}
\left(\frac{m_D^2}{\LU^2}\right)^{\dU-1}~.
\end{eqnarray}
The CP asymmetric parameters $C_{DD}$ and $S_{DD}$ are given by
\begin{eqnarray}
C_{DD} &=& \frac{2\Delta_{DD}}{{\bar f}_{DD}}\sin\phi_w \sin(\dU\pi)~,\\
S_{DD} &=& \frac{-1}{{\bar f}_{DD}}\left[
   \sin 2\beta_0 + 2\Delta_{DD}\cos(\dU\pi)\sin(2\beta_0-\phi_w)
   +\Delta_{DD}^2\sin(2\beta_0-2\phi_w)\right]~,
\end{eqnarray}
where
\begin{equation}
{\bar f}_{DD} = 1+2\Delta_{DD}\cos\phi_w\cos(\dU\pi)+\Delta_{DD}^2~,
\end{equation}
and $\beta_0$ is the relevant CKM phase and $\phi_w$ is the weak phase. 
We fix $\phi_w=3\pi/2$ in this analysis.
For details, see \cite{Lenz}.
Experimental data are \cite{Belle2012_DD}
\begin{eqnarray}
{\rm Br}(B\to D^+D^-) &=& (2.12\pm 0.16\pm 0.18)\times 10^{-4}~,\\
C_{DD} &=&  -0.43\pm 0.16\pm 0.05~,\\
S_{DD} &=& -1.06^{+0.21}_{-0.14}\pm 0.08~.
\end{eqnarray}
%
\par
Now consider $B_s\to\mu\mu$ decay.
The measured branching ratio is
\begin{equation}
 \Br(B_s\to\mu^+\mu^-)=\left(3.2^{+1.5}_{-1.2}\right)\times 10^{-9}~,\label{EXP_Bsmumu}\\
 \end{equation}
Theoretically the branching ratio can be written as
\begin{equation}
 \Br(B_s\to\mu\mu)=\Br_{\SM}\cdot|P|^2~,
\label{Br}
\end{equation}
where $\Br_\SM$ is the SM prediction,
\begin{equation}
 \Br_\SM=(3.23\pm 0.27)\times 10^{-9}~,\label{SMBs}\\
\end{equation}
and
\begin{equation}
 P=1+\frac{m_{B_s}^2}{2m_\mu}\frac{m_b}{m_b+m_s}\frac{C_P}{C_{10}^{\rm SM}}~.
\label{P}
\end{equation}
Here the coefficients $C_{10}^{\rm SM}$ and $C_P$ are given by
\begin{eqnarray}
C_{10}^{\rm SM}&=&
  -\frac{1}{\sin^2\theta_W}\eta_Y Y_0(x_t)~,\label{C10}\\
C_P&=&
  \frac{\sqrt{2}\pi}{G_F\alpha(V_{tb}V_{ts}^*)}\frac{A_\dU e^{i\phi_\calU}}{\sin\dU\pi}
  \left(\frac{m_{B_s}}{\Lambda_\calU}\right)^{2\dU}
  \left(\frac{2m_\mu}{m_{B_s}^4}\right)\left(c_S^q\cdot c_S^\ell\right)^*~,\label{CP}
\end{eqnarray}
where $x_t=m_t^2/m_W^2$, $Y(x)=\eta_Y Y_0(x)$, and
\begin{equation}
 Y_0(x)=\frac{x}{8}\left[\frac{x-4}{x-1}+\frac{3x}{(x-1)^2}\ln x\right]~,~~~
\eta_Y=1.0113~.
\end{equation}
For more details, see \cite{JPL_Bs2mumu}.
%
\par
Finally, the branching ratio of $B_c\to\tau\nu$ decay is
\begin{equation}
{\rm Br}(B_c\to\tau\nu) =
   0.02\left(\frac{f_{B_c}}{0.43~{\rm GeV}}\right)^2
   \Big|1+C_V^{L,\tau}+4.3(C_S^{R,\tau}-C_S^{L,\tau})\Big|^2~.
\end{equation}
Note that large $R(\Ds)$ favors large $C_V^{L,\tau}$ or $\left(C_S^{R,\tau}-C_S^{L,\tau}\right)$,
which makes $\Br(B_c\to\tau\nu)$ larger.
It means that small $\Br(B_c\to\tau\nu)$ would provide severe constraints on model parameters \cite{Alonso2016}.
The predicted upper bound ranges from $10\%$ to $60\%$ \cite{Akeroyd}.
In this analysis we require the branching ratio less than $30\%$.
%
\begin{table}
\begin{tabular}{ccccccc}\hline
$R(D)$ & $R(D^*)$ & $P_\tau(D)$ & $P_\tau(D^*)$ & $F_L(D^*)$ & ${\rm Br}(B_c\to\tau\nu)$ 
& $\chi^2_{\rm min}/{\rm d.o.f.}$ \\\hline\hline
 $0.371$ & $0.266$ & $0.452$ & $-0.445$ & $0.476$ & $7.87\times 10^{-2}$ & $2.22$ \\\hline
\end{tabular}
\caption{Best-fit values.}
\label{T2}
\end{table}
%
\par
For the fitting, we simply put $c_V^{q,\ell}=c_S^{q,\ell}$.
As mentioned before, the vector $\calU$ contribution is highly suppressed.
We found that $C_V^{L,\tau}$ is smaller than $C_S^{L(R),\tau}$ by a few orders.
This is due to the factor of $m_B^2/\LU^2\sim\calO(10^{-6})$ for $\LU\simeq 1~{\rm TeV}$.
Thus the couplings $c_V^{q,\ell}$ must be of order $\calO(10^3)$ in order to compete the scalar contributions.
\par
\begin{figure}
\begin{tabular}{cc}
\hspace{-1cm} \includegraphics[scale=0.12]{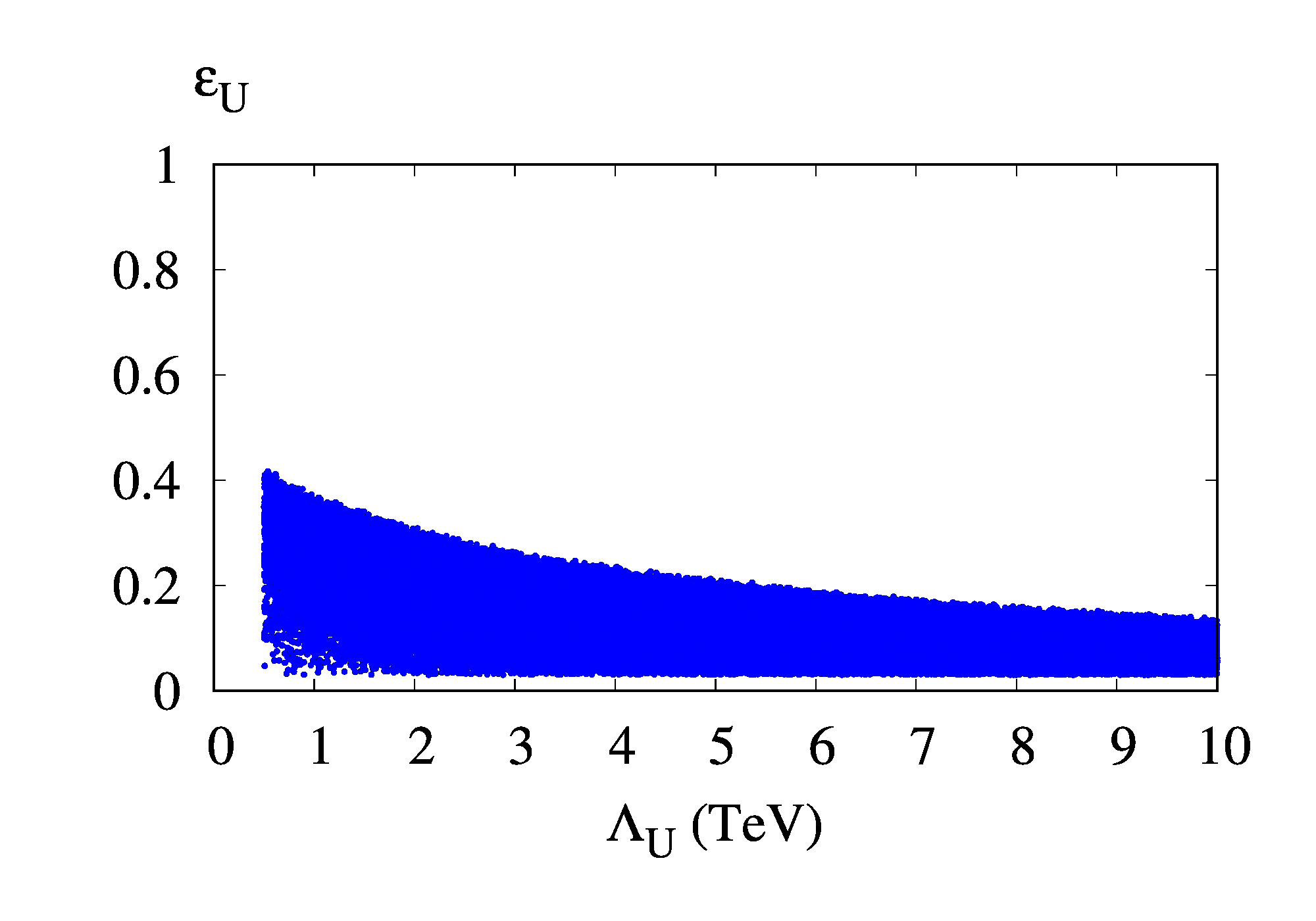} & 
\hspace{-3cm} \includegraphics[scale=0.12]{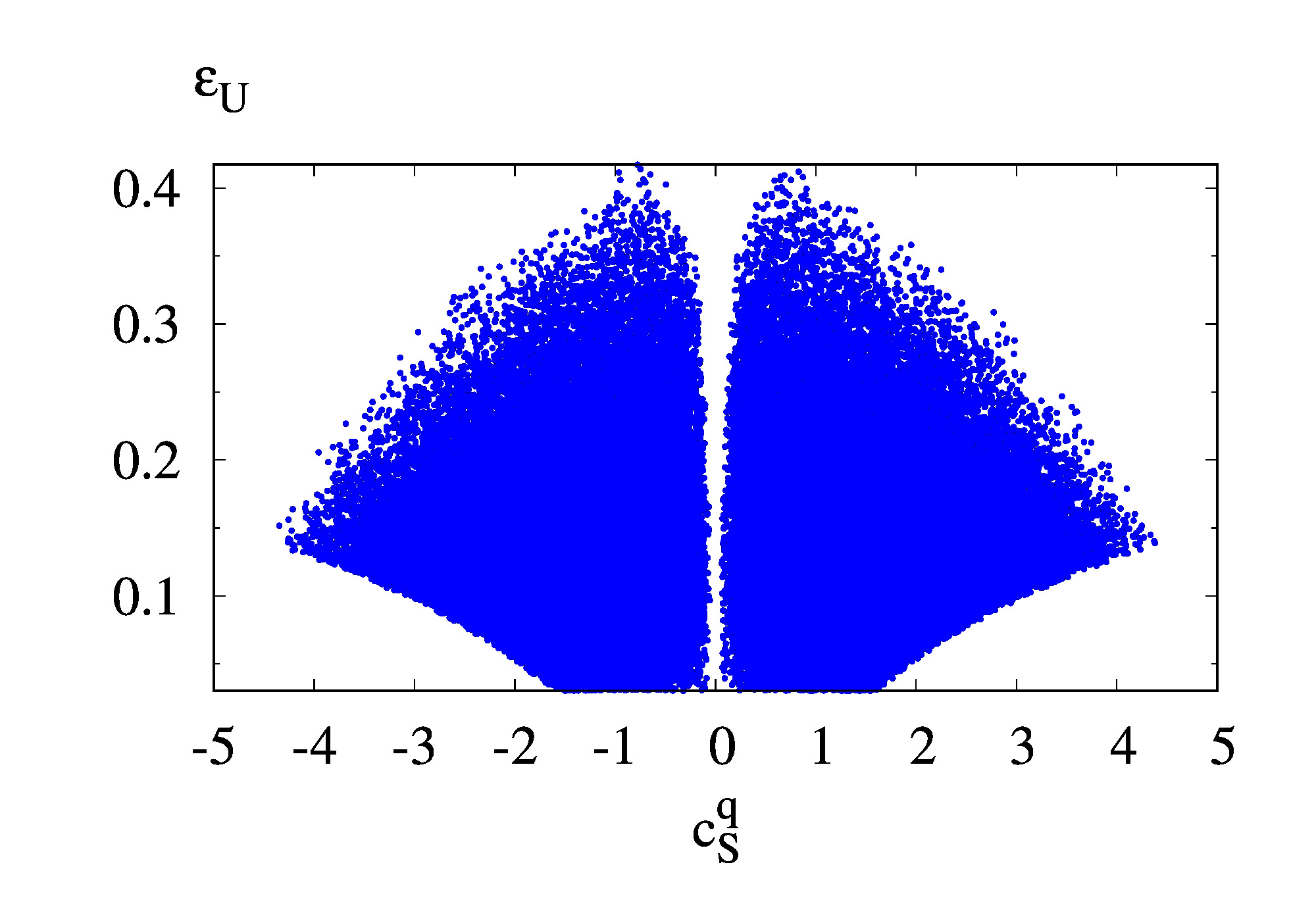}\\
\hspace{-1cm} (a) & \hspace{-3cm} (b) \\
\hspace{-1cm} \includegraphics[scale=0.15]{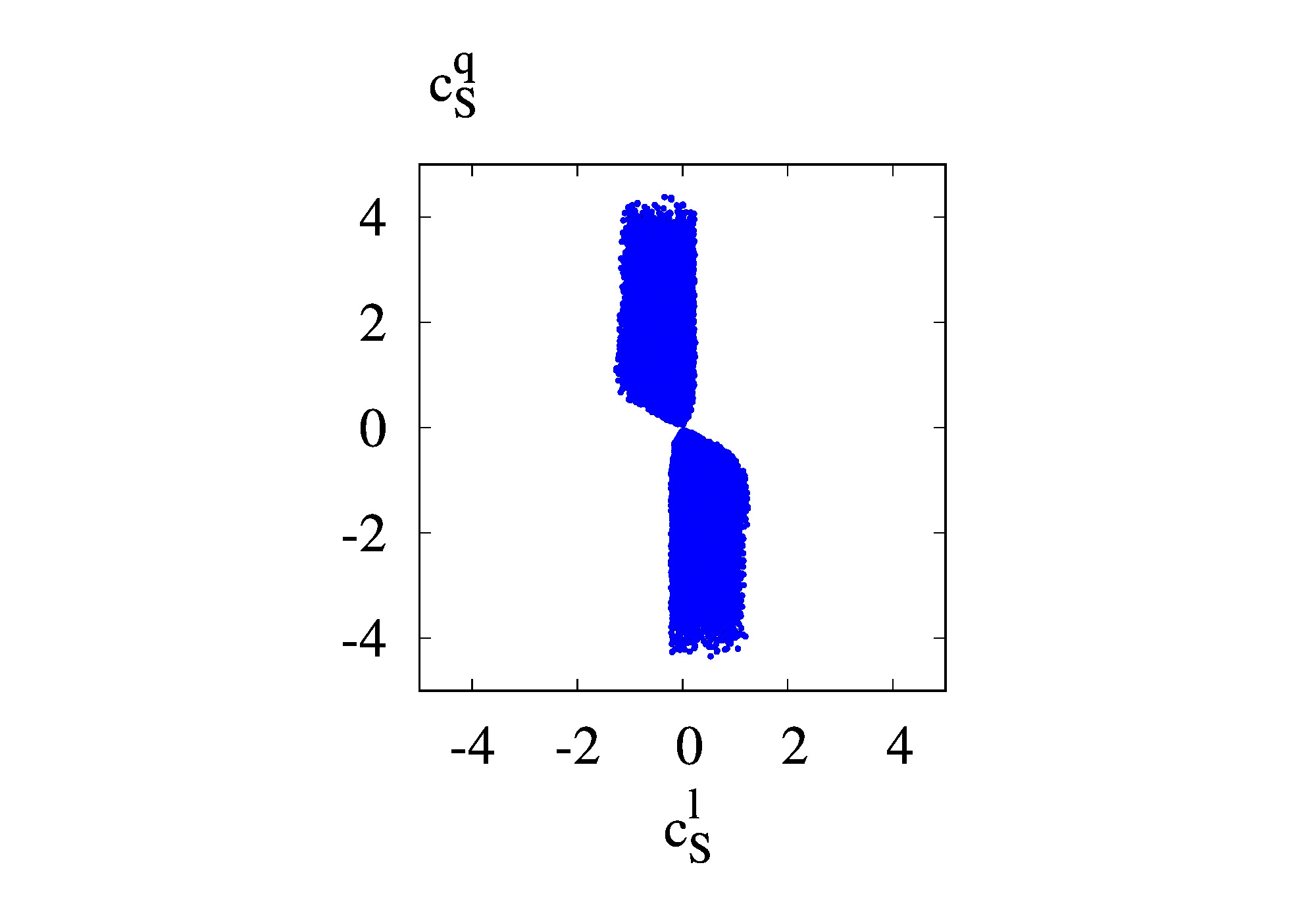} & 
\hspace{-3cm} \includegraphics[scale=0.15]{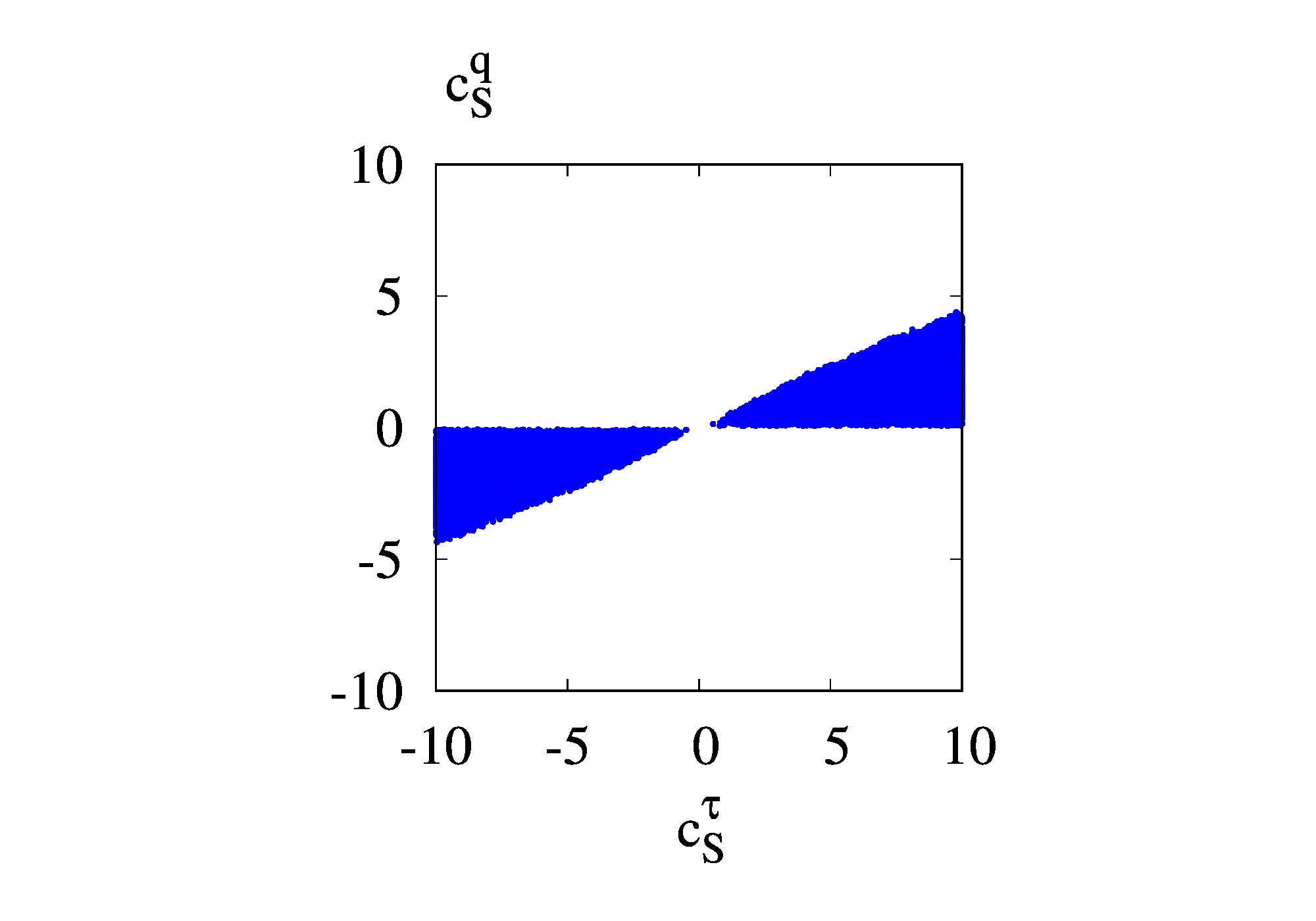}\\
\hspace{-1cm} (c) & \hspace{-3cm} (d) \\
\hspace{-1cm} \includegraphics[scale=0.12]{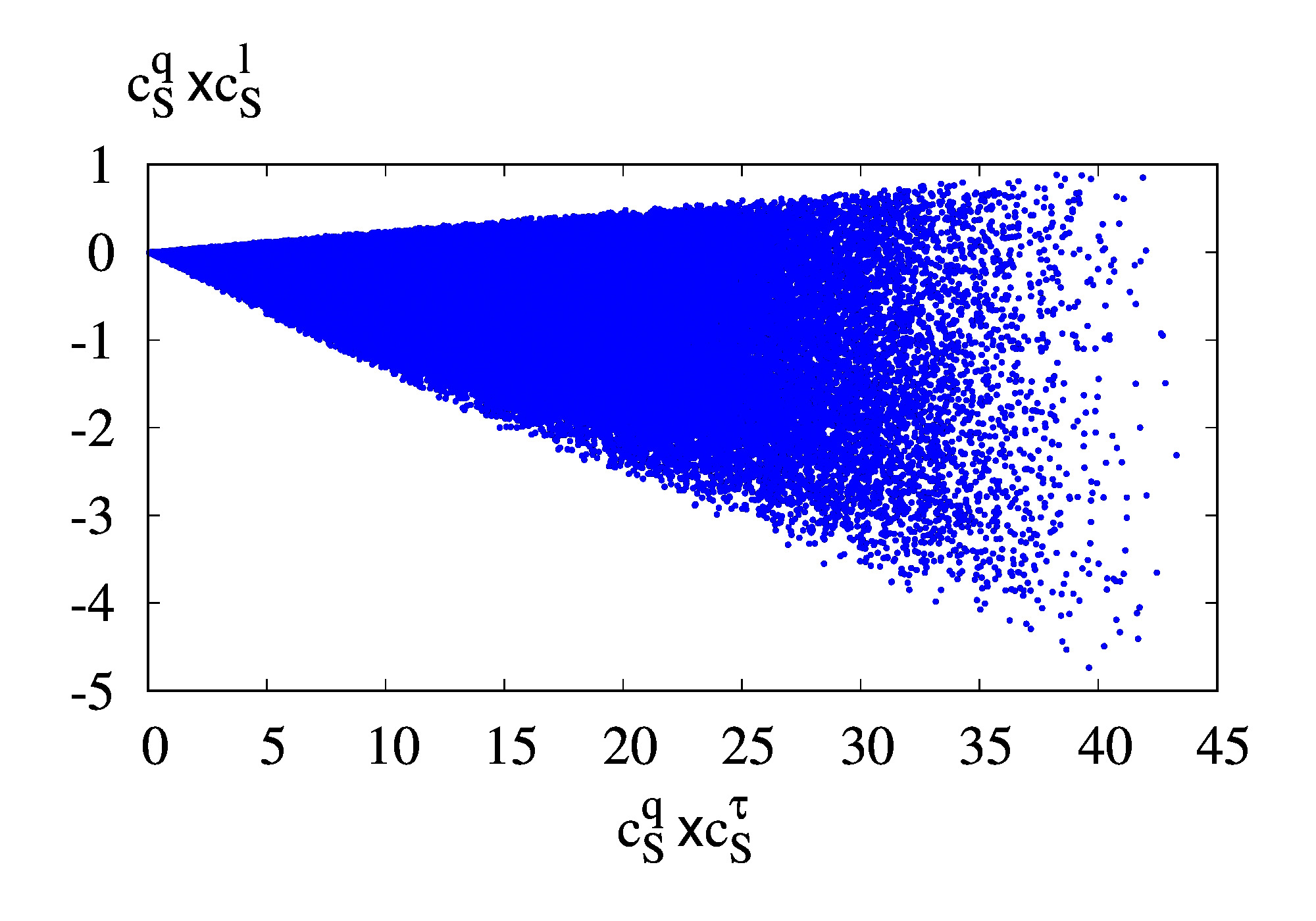}& \\
\hspace{-1cm} (e) & 
\end{tabular}
\caption{\label{paraSP} Parameter spaces allowed at the $2\sigma$ level
for (a) $\eU$ vs $\LU$ (TeV), (b) $\eU$ vs $c_S^q$, 
(c) $c_S^q$ vs $c_S^\ell$, (d) $c_S^q$ vs $c_S^\tau$, 
and (e) $c_S^qc_S^\ell$ vs $c_S^qc_S^\tau$, where $\ell =e,\mu$.}
\end{figure}

Our results are summarized in Table \ref{T2} where the best-fit values for the minimum $\chi^2$ are given.
The best-fit value of $R(D)$ is quite larger than the SM prediction, and that of $\Br(B_c\to\tau\nu)$ is below $10\%$. 
Figure \ref{paraSP} shows the allowed region of model parameters at the $2\sigma$ level.
%
The pattern shown in Fig.\ \ref{paraSP} (a) is typical in unparticle scenario. 
Unparticle contributions appear in the form of $C_\calU\left(m_B^2/\LU^2\right)^\eU$ where
$C_\calU$ is some combination of the relevant couplings.
For large values of $\LU$, large $\eU$ suppresses the new contribution too much to provoke meaningful effects.
The CMS collaboration had put a lower limit on $\LU$ with respect to $\eU$ at high-energy collisions \cite{CMS13TeV}.
The results are for larger values of $c_S^{q,\ell}\sim\calO(10^3)$ where $\LU$ must be large enough to moderate
the unparticle effects.
As discussed in \cite{JPL_Bs2mumu} the results are consistent with ours.
\par
The quark coupling $c_S^q$ in relation with $\eU$ is plotted in Fig.\ \ref{paraSP} (b).
Figure \ref{paraSP} (c) shows the quark coupling vs lepton one. 
For these couplings $B\to D^+D^-$ and $B_s\to\mu^+\mu^-$ decays provide severe bounds on 
$c_S^q$ and $c_S^\ell$, respectively.
$B_s$-${\bar B}_s$ mixing could also constrain the $c_S^q$ coupling, 
but the result is rather weak because uncertainties in the experimental data are still large.
\par
In Fig.\ \ref{paraSP} (d) $c_S^q$ vs $c_S^\tau$ is shown.
Large values of $c_S^\tau$ are favored, as expected.
Note that a small $c_S^q$ multiplied by a small $c_S^\tau$ results in small Wilson coefficients for $R(\Ds)$,
which are disfavored by experimental data.
Figures \ref{paraSP} (c) and \ref{paraSP} (d) show that $c_S^\ell$ favors the opposite sign $c_S^q$ while
$c_S^\tau$ does the same sign $c_S^q$.
Note that the couplings appear in the form of $c_S^qc_S^{\ell,\tau}$ in the Wilson coefficients.
As a result, $c_S^qc_S^\tau$ is positive while $c_S^qc_S^\ell$ is mostly negative, as in Fig.\ \ref{paraSP} (e).
We found that the value of $\chi^2$ gets smaller when $c_S^q c_S^\ell$ approaches to zero.
\par
\begin{figure}
\begin{tabular}{cc}
\hspace{-1cm} \includegraphics[scale=0.13]{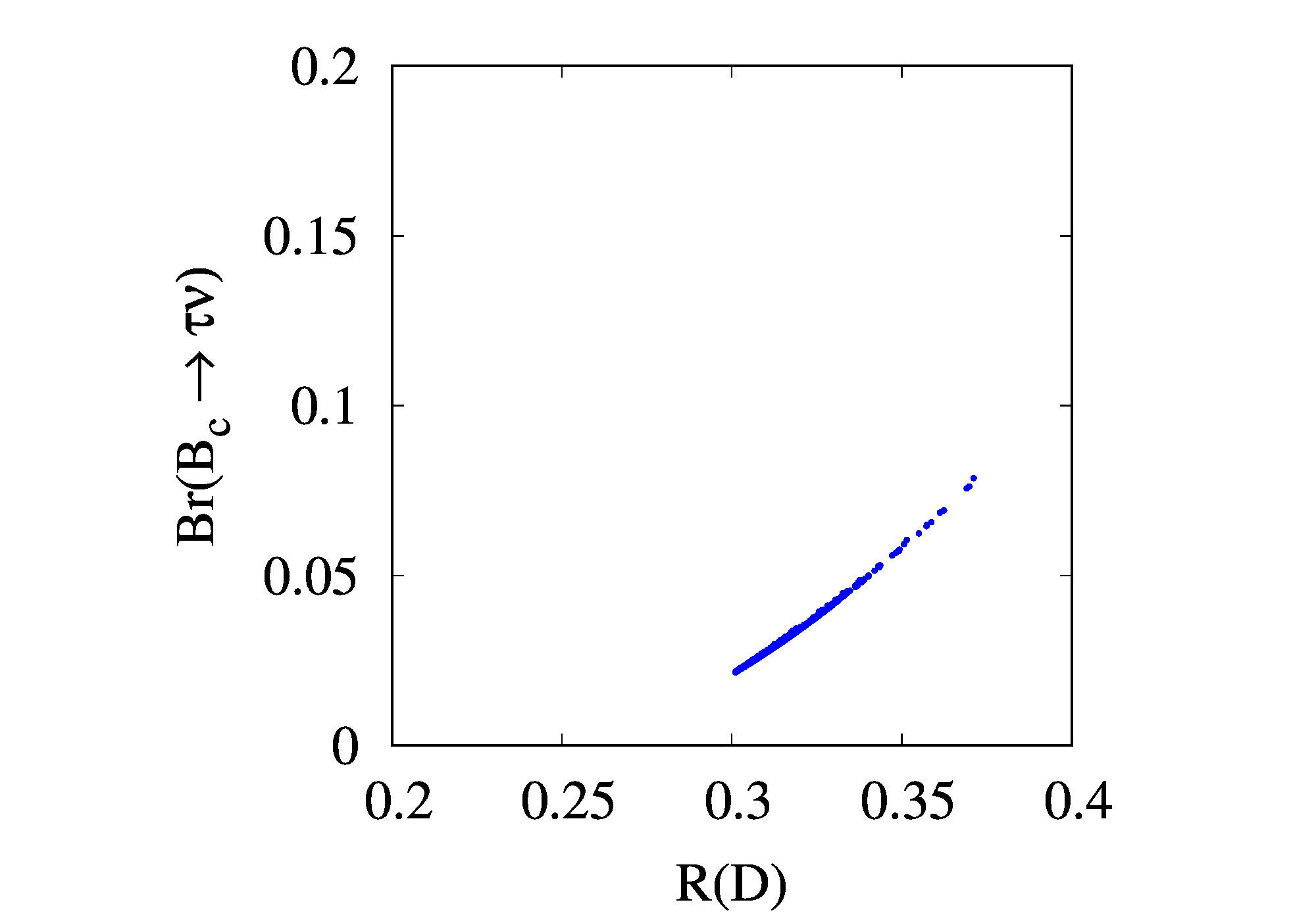} & 
\hspace{-1cm} \includegraphics[scale=0.13]{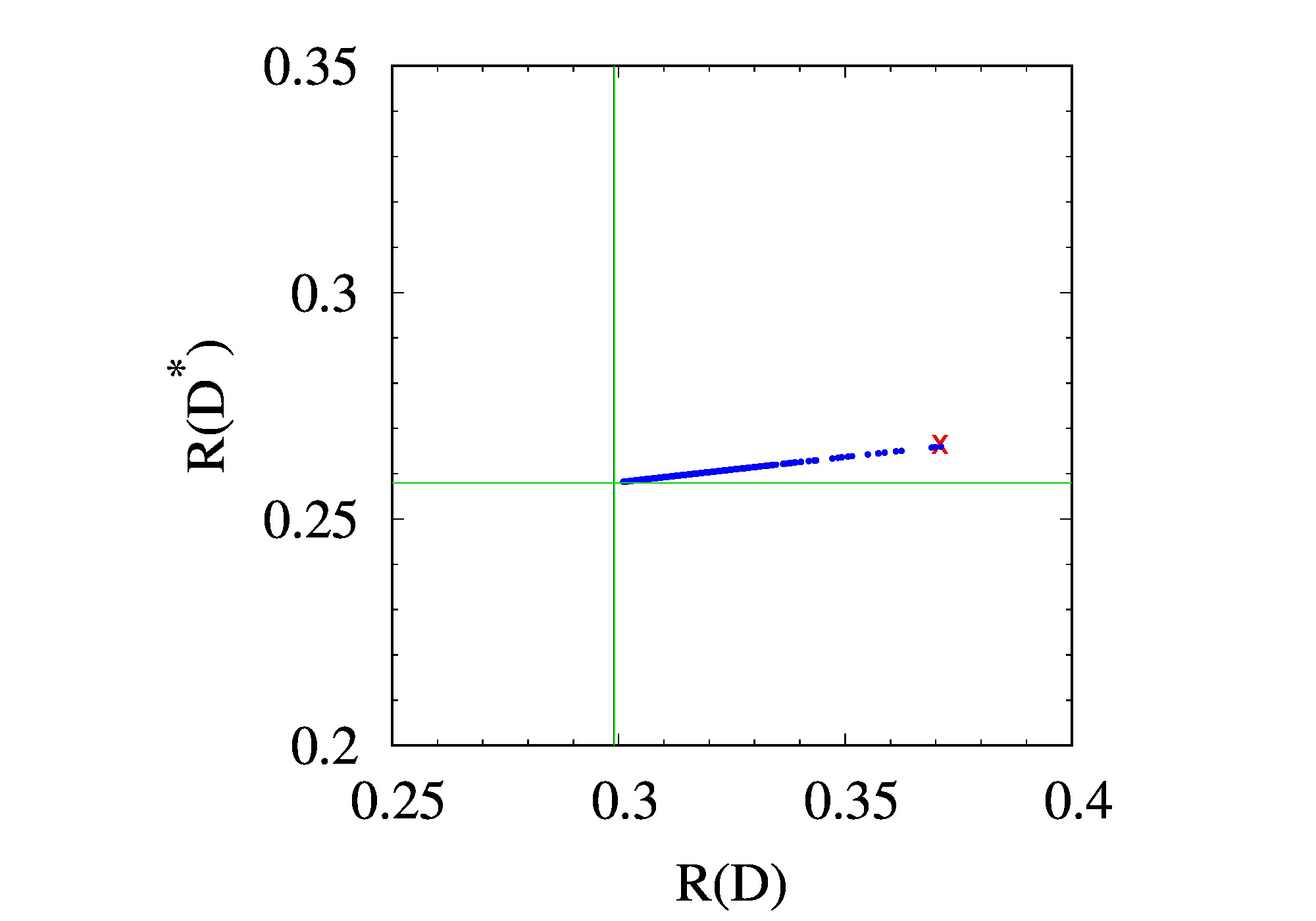}\\
\hspace{-1cm} (a) & \hspace{-1cm} (b) \\
\hspace{-1cm} \includegraphics[scale=0.13]{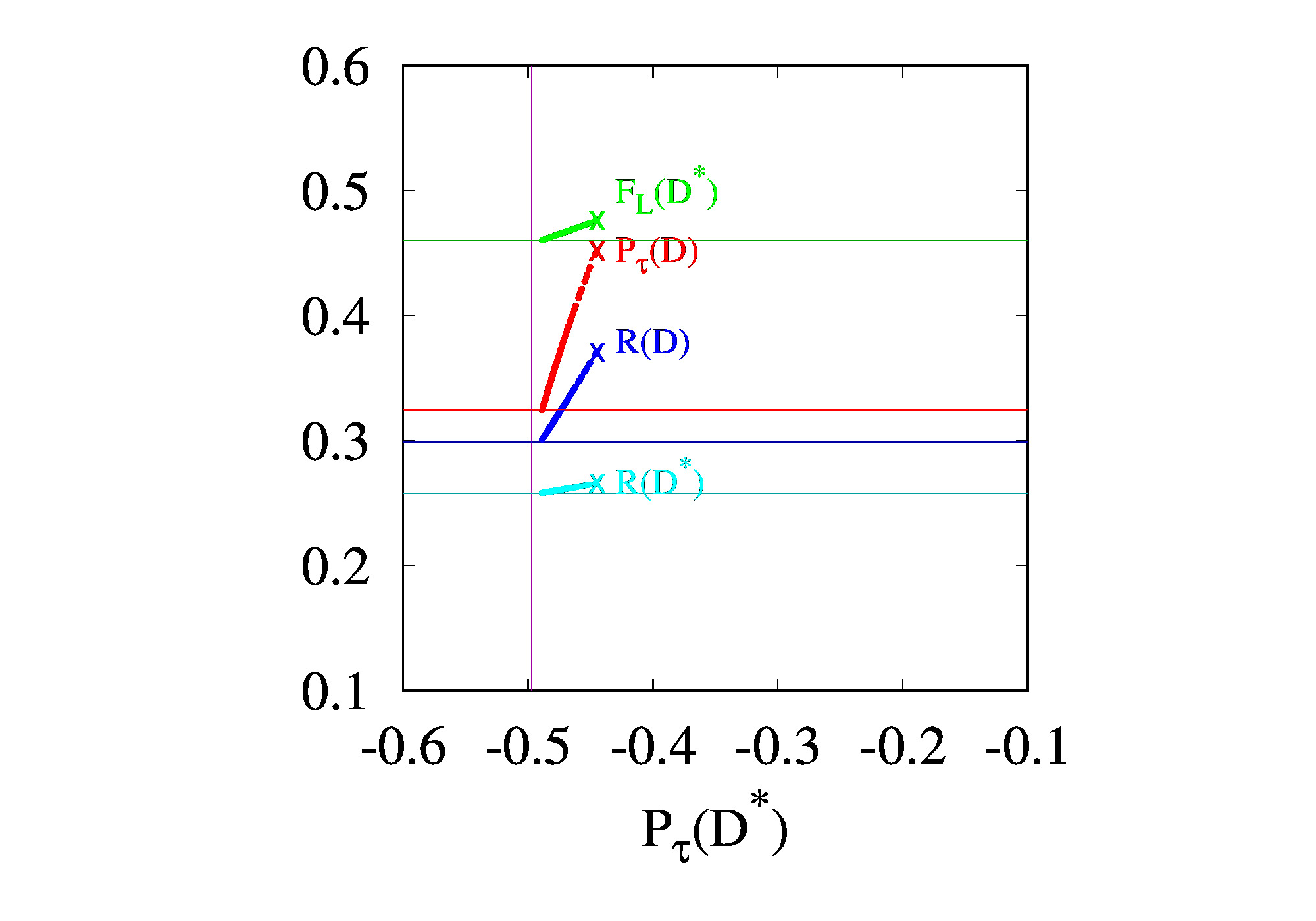} &\\
\hspace{-1cm} (c) &
\end{tabular}
\caption{\label{obs} Predicted ranges of observables at the $2\sigma$ level. 
In (a) $\Br(B_c\to\tau\nu)$ vs $R(D)$; in (b) $R(D^*)$ vs $R(D)$ ;in (c) $R(\Ds)$, $P_\tau(D)$, and $F_L(D)$ vs $P_\tau(D^*)$.
In (b) and (c) vertical and horizontal lines are the SM predictions while the marks "x" denote the best-fit values.}
\end{figure}

On the other hand, we found that $\Br(B_c\to\tau\nu)$ puts no significant bounds on $c_S^\tau$.
We found that requiring $\Br(B_c\to\tau\nu)<30\%$ has almost no difference from $\Br(B_c\to\tau\nu)<10\%$.
All the allowed points satisfy $\Br(B_c\to\tau\nu)<10\%$, as shown in Fig.\ \ref{obs} (a).
\par
Some comments are in order for $b\to s\gamma$, which provides significant constraints on flavor physics. 
For unparticles, $b\to s\gamma$ constrains the quark couplings as studied in \cite{He2008}.
The analysis of \cite{He2008} was done at fixed $\LU=1~{\rm TeV}$.
According to \cite{He2008}, smaller quark couplings are allowed for smaller $\eU$,
and for larger values of $\eU$ the constraints become very weak.
This is due to the suppression factor of $\left(M_{\rm EW}^2/\LU^2\right)^\eU$.
For example, the product of scalar quark couplings $\lambda_{ss}^{YD}\lambda_{bs}^{YD}$ has a bound of
$-0.0026$ at $\eU=0.1$ \cite{He2008}.
In our language, $c_S^q = (v/m_q)\lambda_{qq'}^{YD}$ where $v$ is the Higgs vacuum expectation value.
Thus the bound of \cite{He2008} would not be strong for our analysis.
%
%
\par
Figure \ref{obs} shows the ranges of observables at the $2\sigma$ level. 
Our predictions for $\Br(B_c\to\tau\nu)$ vs $R(D)$ are given in Fig.\ \ref{obs} (a).
Note that keeping $\Br(B_c\to\tau\nu)$ small and making $R(D)$ large enough to fit the data 
is possible in our scenario.
As mentioned before, requiring small branching ratio for $B_c\to\tau\nu$ is not powerful for the model parameters.
We found that the constraint from $B\to D^+D^-$ decay already suppresses $\Br(B_c\to\tau\nu)$ strongly. 
In Figs.\ \ref{obs} (b) and (c), the marks "x" represent the best-fit values and
the straight lines are the central values of the SM predictions.
As seen in Fig.\ \ref{obs} (b), $R(D^*)_\SM$ is at the edge of our allowed range while $R(D)_\SM$ is outside.
And the best-fit values lie in the far side of the region.
In Fig.\ \ref{obs} (c) $R(\Ds)$, $P_\tau(D)$, and $F_L(D^*)$ vs $P_\tau (D^*)$ are provided.
We find that the SM prediction of $P_\tau(D^*)$ lies outside of our allowed region.
%
\section{Conclusions}
In conclusion, we have analyzed the $B$ anomalies in the unparticle scenario.
We included the scalar and vector unparticles, 
and found that vector contributions are a few orders of magnitude smaller than scalar ones.
We implemented the global fit to the relevant observables by minimizing $\chi^2$.
Various constraints are imposed on the model parameters.
Compared to our previous works \cite{JPL_BsBsbar,JPL_Bs2mumu}, 
$\chi^2_{\rm min}/{\rm d.o.f.}$ in this work is slightly larger than before.
It's because $B\to D^+D^-$ decay puts quite a strong constraint on the quark coupling.
As discussed in \cite{JPL_nmUED}, a similar situation could occur in the nmUED model
when the quark sector is severely constrained.
On the other hand, $\Br(B_c\to\tau\nu)$ might not be a strong restriction to the model parameters.
One can safely keep the branching ratio at low values.
This was also true for our previous works \cite{JPL_nmUED,JPL_RD_U}.
We expect more data on anomalous observables could test the unparticle scenario further in near future.

%
%
%

\end{document}